\documentclass[aps,prb,twocolumn,superscriptaddress,showpacs]{revtex4-1}
\usepackage{graphicx}
\usepackage{units}
\usepackage{textcomp}
\usepackage{color}
\usepackage[normalem]{ulem}
\usepackage{mathrsfs}
\usepackage{amsbsy}

\usepackage{epstopdf}

\begin{document}

\title{Ultrafast shift and rectification photocurrents in GaAs quantum wells:
Excitation intensity dependence and the importance of bandmixing}

\author{Huynh Thanh Duc}
\affiliation{Department of Physics and CeOPP, Universit\"at Paderborn, Warburger Str. 100, D-33098 Paderborn, Germany}
\affiliation{Ho Chi Minh City Institute of Physics, Vietnam Academy of Science and Technology, Mac Dinh Chi Str. 1, District 1, Ho Chi Minh City, Vietnam}
\author{Reinold Podzimski}
\affiliation{Department of Physics and CeOPP, Universit\"at Paderborn, Warburger Str. 100, D-33098 Paderborn, Germany}
\author{Shekhar Priyadarshi}
\affiliation{Physikalisch-Technische Bundesanstalt, Bundesallee 100, D 38116 Braunschweig, Germany}
\author{Mark Bieler}
\affiliation{Physikalisch-Technische Bundesanstalt, Bundesallee 100, D 38116 Braunschweig, Germany}
\author{Torsten Meier}
\affiliation{Department of Physics and CeOPP, Universit\"at Paderborn, Warburger Str. 100, D-33098 Paderborn, Germany}

\begin{abstract}
A microscopic approach that is based on the multisubband semiconductor Bloch equations formulated in the basis of a 14-band ${\mathbf k} \cdot {\mathbf p}$ model is employed to compute
the temporal dynamics of photocurrents in GaAs quantum wells following the excitation with femtosecond laser pulses. This approach provides a transparent description of the interband, intersubband, and intraband excitations, fully includes all resonant as well as off-resonant excitations, and treats the light-matter interaction non-perturbatively. For linearly polarized excitations the photocurrents contain contributions from shift and rectification currents. We numerically compute and analyze these currents generated by the excitation with femtosecond laser pulses for [110]- and [111]-oriented GaAs quantum wells.
It is shown that the often employed perturbative $\chi^{(2)}$-approach breaks down for peak fields larger than about 10~kV/cm and that non-perturbative effects lead to a reduction of the peak values of the shift and rectification currents and to temporal oscillations which originate from Rabi flopping.
In particular, we find a complex oscillatory photon energy dependence of the magnitudes of the shift and rectification currents.
Our simulations demonstrate that this dependence is the result of mixing between the heavy- and light-hole valence bands.
This is a surprising finding since the bandmixing has an even larger influence on the strength of the photocurrents than the absorption coefficient.    
For [110]-oriented GaAs quantum wells the calculated photon energy dependence is compared to experimental results and 
a good agreement is obtained which validates our theoretical approach. 
\end{abstract}


\maketitle

\section{Introduction}

In systems of sufficiently reduced symmetry it is possible to generate electrical 
currents on ultrafast time scales without any applied bias
simply by the optical excitation with suitable laser pulses.
In particular, in noncentrosymmetric semiconductor systems
the non-vanishing second-order optical susceptibility $\chi^{(2)}$
allows one to generate photocurrents by excitation with a single
optical pulse.\cite{belinicher,sturman,khurgin,zhang,kral,sipe,bhat00,ganichev00,ganichev02,cote,bhat05,laman,bieler06,bieler06jap,bieler07,n1,golub,nastos,duc10,n2,n2a,n3,n4,priyadarshi} As was shown by Sipe {\it et al.}\cite{sipe},
$\chi^{(2)}$ contains three different contributions which correspond to three
kinds of photocurrent named injection,  shift, and rectification currents.
Injection currents are caused by asymmetric populations of spin-polarized carriers in k-space which originate from Dresselhaus and/or Rashba spin splittings in systems of sufficiently reduced symmetry.
To excite spin-polarized carriers circularly polarized light fields are required.
Without any additional electric or magnetic fields, injection currents do not exist in bulk GaAs, 
they are, however, present in GaAs quantum wells (QW) with lower symmetry.\cite{golub,priyadarshi,duc10} 

Under the action of an optical field that induces interband transitions
the electronic charge density
in the noncentrosymmetric GaAs crystal
is shifted in real space from the As atoms towards the neighboring Ga atoms
and this process leads to the so-called shift current.\cite{nastos} 
Additionally the spatial shift of bound electrons leads to a static polarization, i.e., optical rectification.
If the optical rectification is time varying, e.g., due to the time dependence of the optical pulse envelope,
it induces another kind of current: the rectification current. 
Unlike injection currents
which are created by resonantly excited carriers and shift currents which require 
a resonant excitation in first order and therefore exist for excitation frequencies
above the band gap, rectification currents are present for all photon energies.

Previously we have developed a microscopic approach\cite{duc10} in which we employed the commonly
used 14-band k.p method to obtain the electron bandstructure and the Bloch functions for
GaAs QWs and used the wave functions to formulate the multiband semiconductor Bloch equations (SBE).
We solved the SBE to describe
the optoelectronic response excited by ultrashort laser pulses.
Using this approach we analyzed injection currents generated by circularly polarized
pulses in GaAs QWs and obtained results in quantitative agreement with experiment.\cite{priyadarshi}
When limiting our approach to a perturbative analysis we were
able to reproduce the magnitude and the dynamics of the GaAs shift current obtained in Ref.~\citenum{nastos},
where on the basis of a
full ab-initio band structure
a second-order ($\chi^{(2)}$) analysis of the optical response was performed.

Here, we further extend our approach to describe besides injection and shift currents also rectification currents.
Thus we are able to describe all three photocurrents that exist in the second-order response of noncentrosymmetric
semiconductors\cite{sipe} in an unified way directly in the time domain
and non-perturbatively in the light-matter interaction.
We would like to point out that (unlike to the case of injection currents)
due to the involved non-resonant excitations to higher bands
the rotating wave approximation cannot be applied for
simulations of shift and rectification currents in the framework of Bloch equations
which significantly increases the numerical effort.
We apply our microscopic approach to analyze
shift and rectification currents in GaAs QW systems.
It is shown that for large enough excitation intensities
non-perturbative effects arising from Pauli blocking lead to a reduction of the peak values of the shift and rectification currents
and to intensity-dependent temporal oscillations which originate from Rabi flopping.
The magnitude and direction of the photocurrents depend on the polarization direction of the incident pulse and
the strength of different inter- and intersubband coherences to the currents strongly depends on the
excitation geometry.
Furthermore, we find an oscillatory dependence of the currents on the photon energy, which is also confirmed experimentally.
Our simulations demonstrate that this dependence is caused by valence bandmixing.
This is an interesting finding since is emphasizes the tremendous influence of bandmixing on transport phenomena
and might be at the origin of new applications.
To keep the numerical requirements within reasonable limits,
excitonic effects are neglected in our present calculations. 

This paper is organized as follows.
In Sec.~II, we present our theoretical approach.
In Sec.~III several numerical results for shift and rectification currents in [110]- and [111]-oriented QWs are discussed.
Experimental results and a comparison between experiment and theory are shown in Sec.~IV. 
Finally, the main results are briefly summarized in Sec.~V.

\section{Theoretical approach}

The energies and wave functions of electrons in a semiconductor QW are described by the Schr\"odinger equation
\begin{eqnarray}\label{schro}
\left[ \frac{\hbar^2}{2m_0}{\mathbf\nabla}^2 + V_0 + H_{\mathrm{SO}} + V_{\mathrm{conf}}\right] \psi = \varepsilon\psi,
\end{eqnarray}
where $V_0$ is the periodic lattice potential, $H_{\mathrm{SO}}$ is the spin-orbit interaction, and $V_{\mathrm{conf}}$ is the confinement potential. To obtain a realistic electronic band structure and
wave functions
near the $\Gamma$-point
we employ a 14-band ${\mathbf k} \cdot {\mathbf p}$ method \cite{rossler,winkler03,elkurdi} within the envelope function approximation.
By choosing the $z$-axis as the growth direction of the QW and writing the electron wave function as $\psi=e^{i\mathbf k_\parallel\cdot\mathbf r_\parallel}\sum\limits_{n=1}^{14} f^n_{\mathbf k_\parallel}(z)u_n$, where ${\mathbf k}_\parallel=(k_x,k_y)$ is the in-plane wave vector, $u_n$ are band-edge Bloch functions, and $f^n_{\mathbf k_\parallel}$ are slowly varying envelope functions Eq.~(\ref{schro}) becomes
\begin{eqnarray}\label{em}
\sum_{m=1}^{14}\left[H_{nm}^{k.p}(\hat\mathbf{k})+V_n(z)\delta_{nm}\right]f^{m}_{\mathbf k_\parallel}(z) =\varepsilon_{\mathbf k_\parallel} f^{n}_{\mathbf k_\parallel}(z),
\end{eqnarray}
where $H^{k.p}_{nm}$ are the matrix elements of the $14$-band ${\mathbf k} \cdot {\mathbf p}$ Hamiltonian with $\hat\mathbf{k}=(k_x,k_y,-i\partial/\partial z)$ and $V_n(z)$ is the band-offset potential of the well
which is taken to be centered at $z=0$. To solve Eq.~(\ref{em}) we expand the envelope function into plane 
waves\cite{elkurdi} $f^n_{{\mathbf k}_\parallel}(z)=\sum\limits_{j=1}^N c^j_{n{\mathbf k}_\parallel}\phi_j(z)$, where $\phi_j(z)=\sqrt{\frac{2}{L}}\sin\left[\frac{\pi j}{L}(z+\frac{L}{2})\right]$ for $-L/2\leq z\leq L/2$ and $\phi_j(z)=0$ for otherwise. This leads to a $14N\times 14N$ eigenvalue equation. By numerical diagonalizations 
of the matrix for several values of ${\mathbf k_\parallel}$ the electronic band structure is obtained.
The number of plane wave functions $N$ and the width $L$  are chosen to
ensure convergence of the results.

To analyze the dynamics of photoexcited semiconductor QWs
we use a Hamiltonian that contains the band structure
and the light-matter interaction in velocity gauge
\begin{eqnarray}\label{ham}
H&=&\sum_{\lambda,\mathbf k_\parallel} \varepsilon_{\lambda\mathbf k_\parallel} a^{\dagger}_{\lambda\mathbf k_\parallel} a^{}_{\lambda\mathbf k_\parallel}\nonumber\\ 
&+& e{\mathbf A}(t)\cdot\sum_{\lambda,\lambda',\mathbf k_\parallel}{\mathbf v}_{\lambda\lambda'\mathbf k_\parallel}a^{\dagger}_{\lambda\mathbf k_\parallel} a^{}_{\lambda'\mathbf k_\parallel},
\end{eqnarray}
where $\varepsilon_{\lambda\mathbf k_\parallel} $ is the energy and $a^{\dagger}_{\lambda\mathbf k_\parallel}$ ($a^{}_{\lambda\mathbf k_\parallel}$) is the creation (annihilation) operator of an electron in
band $\lambda$ with wave vector ${\mathbf k}_\parallel$, ${\mathbf A}(t)$ is the vector potential of the light field, and ${\mathbf v}_{\lambda\lambda'\mathbf k_\parallel}$ is the velocity matrix element between Bloch states
\begin{eqnarray}
{\mathbf v}_{\lambda\lambda'{\mathbf k}_\parallel}=\frac{1}{\hbar}\sum_{n,m=1}^{14}\int dz\ f^{n *}_{\lambda{\mathbf k}_\parallel} ({\mathbf\nabla}_{\mathbf k}H^{k.p})_{nm} f^{m}_{\lambda'{\mathbf k}_\parallel}.
\end{eqnarray}
From the Heisenberg equation of motion we obtain the well-known multiband SBE\cite{haug,pasenow,duc10} which describe the time evolution of the microscopic interband
and intersubband polarizations $p_{\lambda\lambda'\mathbf k_\parallel}=\langle a^{\dagger}_{\lambda\mathbf k_\parallel} a^{}_{\lambda'\mathbf k_\parallel} \rangle$ with $\lambda\neq\lambda'$ and the electron occupations $n_{\lambda\mathbf k_\parallel}=\langle a^{\dagger}_{\lambda\mathbf k_\parallel} a^{}_{\lambda\mathbf k_\parallel} \rangle$
\begin{widetext}
\begin{eqnarray}
\frac{\partial }{\partial t}p^{}_{\lambda\lambda'{\mathbf k}_\parallel} (t) &=&i(\omega_{\lambda\lambda'{\mathbf k}_\parallel}+i/\tau_2)\ p^{}_{\lambda\lambda'{\mathbf k}_\parallel}
+\frac{i}{\hbar}e{\mathbf A}(t)\cdot{\mathbf v}^{}_{\lambda'\lambda{\mathbf k}_\parallel}(n^{}_{\lambda'{\mathbf k}_\parallel}-n^{}_{\lambda{\mathbf k}_\parallel})\nonumber\\
&+&\frac{i}{\hbar}e{\mathbf A}(t)\cdot\left[\sum_{\nu\neq\lambda'}{\mathbf v}^{}_{\nu\lambda{\mathbf k}_\parallel} p^{}_{\nu\lambda'{\mathbf k}_\parallel} -\sum_{\nu\neq\lambda}{\mathbf v}^{}_{\lambda'\nu{\mathbf k}_\parallel} p^{}_{\lambda\nu{\mathbf k}_\parallel} \right],\label{SBE3}
\end{eqnarray}
\begin{eqnarray}
\frac{\partial }{\partial t}n^{}_{\lambda{\mathbf k}_\parallel} (t) =-\frac{2}{\hbar}\ \mathrm{Im}\left[ e{\mathbf A}(t)\cdot\sum_{\lambda'\neq\lambda}{\mathbf v}^{}_{\lambda'\lambda{\mathbf k}_\parallel}p^{}_{\lambda'\lambda{\mathbf k}_\parallel}\right] -\frac{1}{\tau_1}\left(n^{}_{\lambda{\mathbf k}_\parallel}-n^{\mathrm{eq}}_{\lambda{\mathbf k}_\parallel}(n, T)\right),\label{SBE1}
\end{eqnarray}
\end{widetext}
where $\omega_{\lambda\lambda'{\mathbf k}_\parallel}=(\varepsilon_{\lambda{\mathbf k}_\parallel}-\varepsilon_{\lambda'{\mathbf k}_\parallel})/\hbar$ is the transition frequency.
$\tau_1$ and $\tau_2$ are introduced phenomenologically in order to model the relaxation of occupations
to quasi-equilibrium and the dephasing of polarizations due to carrier-phonon and carrier-carrier scattering.
In the numerical calculations, we use typical values of $\tau_1=120$~fs and $\tau_2=82.3$~fs.\cite{duc05,pasenow}
$n^{\mathrm{eq}}_{\lambda{\mathbf k}_\parallel}(n,T)$ is a
quasi-equilibrium thermal occupation at $T=300$~K which has the same carrier density $n(t)$
as the actual time-dependent occupation.

The exciting light field with a central frequency $\omega$ is given by
\begin{eqnarray}
{\mathbf E}(t)={\mathbf E}_\omega(t) e^{i\omega t} + c.c.,
\end{eqnarray}
where ${\mathbf E}_\omega$ is the slowly-varying envelope function.
To describe Gaussian-shaped pulses
which propagate along the $z$-direction
we use ${\mathbf E}_\omega$ in the form
\begin{eqnarray}
{\mathbf E}_\omega(t)=E_0e^{-\frac{t^2}{2\tau_L^2}} (\cos\theta,\sin\theta e^{i\phi},0),
\end{eqnarray}
where $E_0$ is the maximal amplitude, $\tau_L$ is the duration of the Gaussian envelope, $\theta$ is the
polarization
angle with respect to the $x$-axis, and $\phi$ is the  phase difference between the $x$- and the
$y$-components of the field. If $\phi=\pm\pi/2$ and $\theta=\pi/4$ the light field is circularly
polarized, whereas $\phi=0$,
which is used in all simulations presented below, corresponds to linearly polarized pulses. 
The vector potential used in the SBE (\ref{SBE3}) and (\ref{SBE1}) is given by
$\mathbf A(t)=-\int\limits_{-\infty}^t {\mathbf E}(t')dt'$.

\begin{figure}
\centering
\includegraphics[width=6cm]{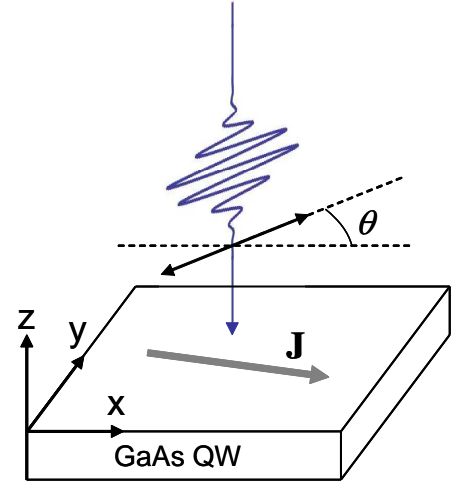}
\caption{(color online) Schematic illustration of the shift and rectification current generation
using linearly polarized light pulses.
$\theta$ is the light polarization angle with respect to the $x$-axis.}
\label{geometry}
\end{figure}

By solving the SBE for an initially unexcited QW system we obtain the time-dependent
occupations $n^{}_{\lambda{\mathbf k}_\parallel}$
and microscopic polarizations $p^{}_{\lambda\lambda'{\mathbf k}_\parallel}$. 
From the occupations we can evaluate the intraband charge current known
as injection photocurrent  via
\begin{eqnarray}
{\mathbf j}^{}_\mathrm{inject}(t)&=&e\sum_{\lambda,{\mathbf k}_\parallel}{\mathbf v}^{}_{\lambda\lambda{\mathbf k}_\parallel} n^{}_{\lambda{\mathbf k}_\parallel} \nonumber\\
&=&e\sum_{c,{\mathbf k}_\parallel}{\mathbf v}^{}_{cc{\mathbf k}_\parallel} n^{}_{c{\mathbf k}_\parallel} + e\sum_{v,{\mathbf k}_\parallel}{\mathbf v}^{}_{vv{\mathbf k}_\parallel} n^{}_{v{\mathbf k}_\parallel},
\end{eqnarray}
where $c$ ($v$) is a band index of the conduction (valence) band. 

From the microscopic polarizations it is possible to compute
the charge current induced by interband and intersubband transitions
\begin{eqnarray}
{\mathbf j}^{}_\mathrm{inter}(t)&=&e\sum_{\lambda,\lambda'\neq\lambda,{\mathbf k}_\parallel}{\mathbf v}^{}_{\lambda\lambda'{\mathbf k}_\parallel} p^{}_{\lambda\lambda'{\mathbf k}_\parallel} \nonumber\\
&=& e\sum_{c,c'\neq c,{\mathbf k}_\parallel}{\mathbf v}^{}_{cc'{\mathbf k}_\parallel} p^{}_{cc'{\mathbf k}_\parallel}
+ e\sum_{v,v'\neq v,{\mathbf k}_\parallel}{\mathbf v}^{}_{vv'{\mathbf k}_\parallel} p^{}_{vv'{\mathbf k}_\parallel}\nonumber\\
&+&  e\sum_{c,v,{\mathbf k}_\parallel} \left( {\mathbf v}^{}_{cv{\mathbf k}_\parallel} p^{}_{cv{\mathbf k}_\parallel} + c.c.\right)
\label{interbandcurrent}
\end{eqnarray} 
as well as the interband and intersubband polarization
\begin{eqnarray}
{\mathbf P}^{}_\mathrm{inter}(t)&=& e\sum_{\lambda,\lambda'\neq\lambda,{\mathbf k}_\parallel}{\mathbf r}^{}_{\lambda\lambda'{\mathbf k}_\parallel} p^{}_{\lambda\lambda'{\mathbf k}_\parallel}\nonumber\\
&=& e\sum_{c,c'\neq c,{\mathbf k}_\parallel}{\mathbf r}^{}_{cc'{\mathbf k}_\parallel} p^{}_{cc'{\mathbf k}_\parallel}
+ e\sum_{v,v'\neq v,{\mathbf k}_\parallel}{\mathbf r}^{}_{vv'{\mathbf k}_\parallel} p^{}_{vv'{\mathbf k}_\parallel} \nonumber\\
&+& e\sum_{c,v,{\mathbf k}_\parallel} \left({\mathbf r}^{}_{cv{\mathbf k}_\parallel} p^{}_{cv{\mathbf k}_\parallel} + c.c.\right),
\label{interbandpolarization}
\end{eqnarray}
where $e{\mathbf r}^{}_{\lambda\lambda'{\mathbf k}_\parallel}$ is the transition dipole moment which can be obtained from the velocity operator via the relation ${\mathbf v}^{}_{\lambda\lambda'{\mathbf k}_\parallel}=i\omega^{}_{\lambda\lambda'{\mathbf k}_\parallel}{\mathbf r}^{}_{\lambda\lambda'{\mathbf k}_\parallel}$. 

The
shift current ${\mathbf j}_\mathrm{shift}$ is defined as the zero-frequency contribution
to the interband photocurrent ${\mathbf j}_\mathrm{inter}$.
Similarly, the optical rectification ${\mathbf P}^{}_\mathrm{rect}$ is given by the slowly varying part of the interband polarization ${\mathbf P}^{}_\mathrm{inter}$.
The time-derivative of the optical rectification provides the rectification current 
\begin{eqnarray}
{\mathbf j}^{}_\mathrm{rect}(t)=\frac{\partial}{\partial t}{\mathbf P}^{}_\mathrm{rect}(t)
\end{eqnarray}
As explained in Sec.~III in more detail,
to obtain ${\mathbf j}_\mathrm{shift}(t)$ and ${\mathbf j}^{}_\mathrm{rect}(t)$ we Fourier transform
${\mathbf j}_\mathrm{inter}(t)$ and $\frac{\partial}{\partial t}{\mathbf P}^{}_\mathrm{inter}(t)$,
respectively,
into the frequency domain, apply a filter around $\omega=0$, and transform back to the time domain.

\begin{figure*}
\begin{center}
\includegraphics[width=16cm]{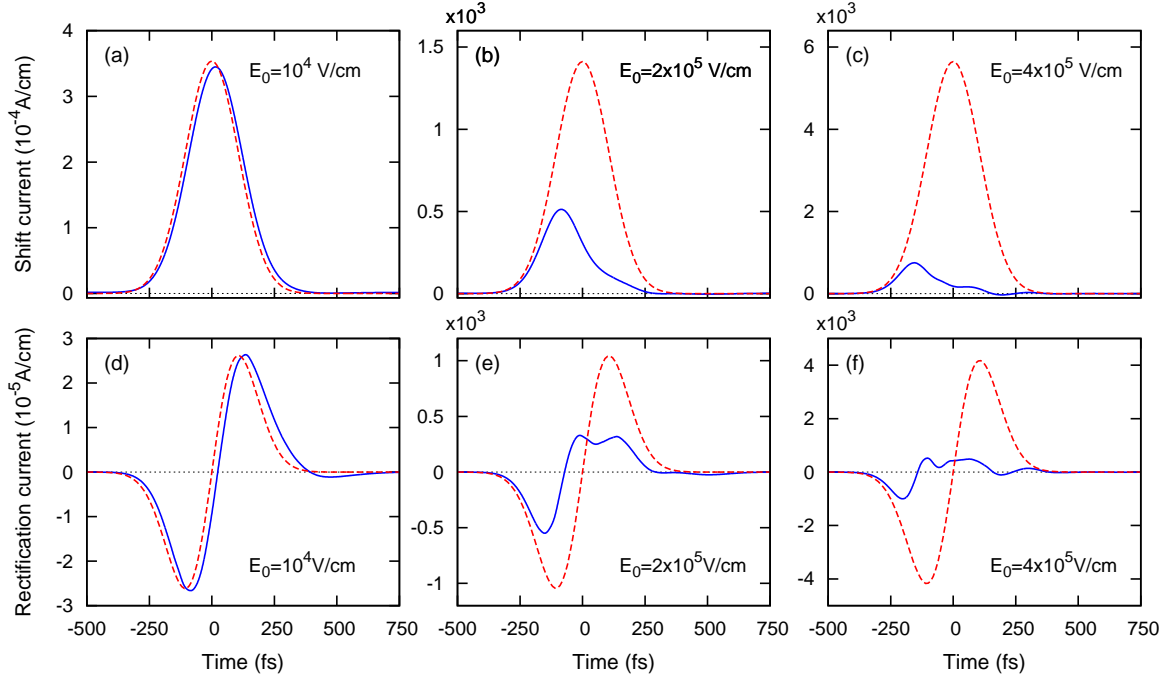}
\caption{(color online) Dynamics of shift and rectification currents in a 10~nm wide [110]-oriented GaAs QW for different peak amplitudes of the laser pulses. The dashed lines are obtained by a perturbative second-order analysis of the light-matter interaction whereas the solid lines are obtained by solving the full SBE.}
\label{t-evol}
\end{center}
\end{figure*}

In the limit of weak excitation intensities the light-matter interaction can be considered as a perturbation.
Using a second-order perturbation expansion of the SBE (\ref{SBE3}) and (\ref{SBE1}), by
applying the rotating wave approximation for the first-order
microscopic polarization $p^{(1)}_{\lambda\lambda'}$,
and neglecting
$2\omega$ terms in the second-order microscopic polarization $p^{(2)}_{\lambda\lambda'}$
we analytically derive
approximate expressions for the injection current, the shift current,
and the optical rectification, respectively, that read~\cite{belinicher,sipe}
\begin{widetext}
\begin{eqnarray}
\frac{\partial}{\partial t}{\mathbf j}^{(2)}_\mathrm{inject}=\frac{2 e^3}{\hbar^2\omega^2}\sum_{c,v,{\mathbf k}_\parallel}({\mathbf v}_{cc{\mathbf k}_\parallel}-{\mathbf v}_{vv{\mathbf k}_\parallel})\left|{\mathbf E}_\omega\cdot{\mathbf v}^{}_{cv{\mathbf k}_\parallel}\right|^2\frac{1/\tau_2}{(\omega-\omega^{}_{cv{\mathbf k}_\parallel})^2+1/\tau_2^2}-{\mathbf j}^{(2)}_\mathrm{inject}/\tau_1,\label{inject2}
\end{eqnarray}
\begin{eqnarray}
{\mathbf j}^{(2)}_\mathrm{shift} =-\frac{2e^3}{\hbar^2\omega^2}\mathrm{Im}\sum_{c,v,{\mathbf k}_\parallel} \frac{{\mathbf E_\omega}\cdot {\mathbf v}^{}_{cv{\mathbf k}_\parallel}}{\omega-\omega^{}_{cv{\mathbf k}_\parallel}+i/\tau_2} 
\left[\sum_{\lambda\neq v} {\mathbf r}^{}_{v\lambda{\mathbf k}_\parallel} ({\mathbf E}^*_\omega \cdot {\mathbf v}^{}_{\lambda c{\mathbf k}_\parallel})- \sum_{\lambda\neq c}({\mathbf E}^*_\omega\cdot{\mathbf v}^{}_{v\lambda{\mathbf k}_\parallel}){\mathbf r}^{}_{\lambda c{\mathbf k}_\parallel} \right],\label{shift2} 
\end{eqnarray}
and
\begin{eqnarray}\label{or}
{\mathbf P}^{(2)}_\mathrm{rect} =\frac{2e^3}{\hbar^2\omega^2}\mathrm{Re}\sum_{c,v,{\mathbf k}_\parallel} \frac{{\mathbf E_\omega}\cdot {\mathbf v}^{}_{cv{\mathbf k}_\parallel}}{\omega-\omega^{}_{cv{\mathbf k}_\parallel}+i/\tau_2} 
\left[\sum_{\lambda\neq v} \frac{{\mathbf r}^{}_{v\lambda{\mathbf k}_\parallel}}{\omega_{v\lambda{\mathbf k}_\parallel}} ({\mathbf E}^*_\omega \cdot {\mathbf v}^{}_{\lambda c{\mathbf k}_\parallel})- \sum_{\lambda\neq c}({\mathbf E}^*_\omega\cdot{\mathbf v}^{}_{v\lambda{\mathbf k}_\parallel}) \frac{{\mathbf r}^{}_{\lambda c{\mathbf k}_\parallel}}{\omega_{\lambda c{\mathbf k}_\parallel}} \right].\label{rect2}
\end{eqnarray}
\end{widetext}

\section{Numerical results}

In this section, we present and discuss results obtained from numerical solutions of the SBE 
for GaAs/Al$_{0.35}$Ga$_{0.65}$As QWs grown in the crystallographic [110]- and [111]-directions.
The band parameters for GaAs and Al$_x$Ga$_{1-x}$As are taken from Ref.~\citenum{winkler03}
and the temperature dependence of the band gap is described by the
Varshni relation. In our calculations we consider room temperature ($T=300$~K).
For the optical excitation we use linearly polarized laser pulses which propagate perpendicular to the 
QW plane, i.e., in $z$-direction, and have a Gaussian envelope with a duration of $\tau^{}_L=150$~fs.
The geometry of the excitation is illustrated in Fig.~\ref{geometry}. 

The band structure of the QW system is obtained from a matrix diagonalization using $14 \, N$ bands,
where we use $N\geq 20$.
Due to this large number of bands, the direct evaluation of the multisubband SBE (\ref{SBE1}) and (\ref{SBE3})
beyond the rotating wave approximation is numerically very intensive.
In order to keep the calculation feasible we limit the number of bands included in the numerics to $40$.
In particular we take into account the eighteen energetically highest valence subbands,
six energetically lowest $s$-like conduction subbands, and sixteen energetically lowest $p$-like conduction subbands. 
Though the optical excitation to $p$-like conduction bands is well nonresonant the presence of these bands in the SBE is necessary.
In second-order processes, the $p$-like conduction band states play the role of intermediate states for the transition of electrons
from the valence to the conduction band and thereby contributions to the shift and rectification currents.

Obtaining shift and rectification currents from numerical solutions of the SBE involves basically the following steps:
(i) solve the SBE, (ii) compute the current ${\mathbf j}_{\mathrm{inter}}(t)$ and the interband polarization ${\mathbf P}_{\mathrm{inter}}(t)$, (iii) Fourier transform ${\mathbf j}_{\mathrm{inter}}(t)$ and ${\mathbf P}_{\mathrm{inter}}(t)$ to the frequency domain and apply a frequency filter to remove the high-frequency components,
(iv) inversely Fourier transform back to time domain
to obtain the shift current ${\mathbf j}_{\mathrm{shift}}(t)$ and the
optical rectification ${\mathbf P}_\mathrm{rect}(t)$.
We carefully checked the convergence versus the number of bands and also compared the
complete numerical solutions with the analytical second-order
approximation, i.e., Eqs.~(\ref{shift2}) and (\ref{rect2}),
where we included all $14 \, N$ bands
to ensure that our approache works properly within excitation conditions considered in the paper.

\begin{figure}
\centering
\includegraphics[width=8cm]{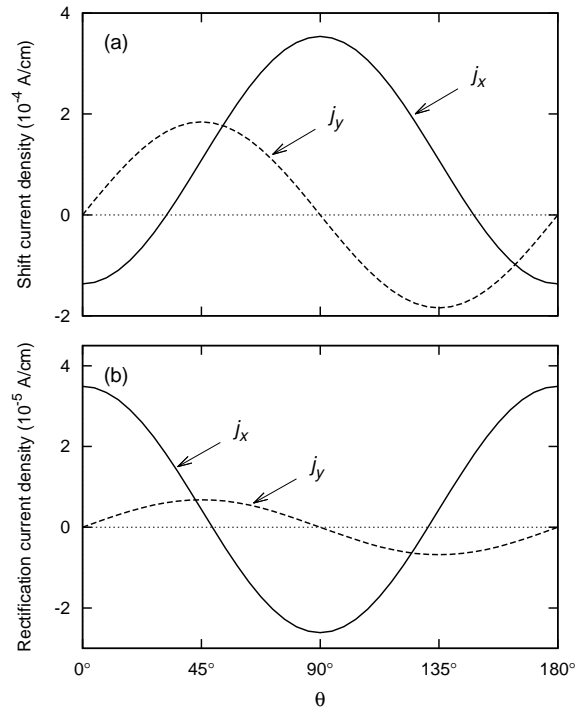}
\caption{Components in $x$- and $y$-directions of the peak value of shift (a) and rectification (b) current densities as function of the light polarization angle $\theta$ in a 10 nm [110]-oriented GaAs QW.}
\label{figure3}
\end{figure}

\begin{figure}
\centering
\includegraphics[width=8cm]{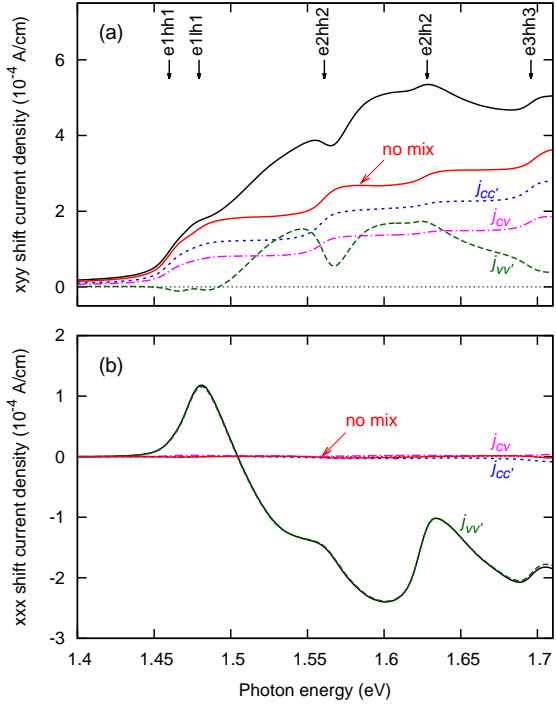}
\caption{(color online) Peak values of the $xyy$ (a) and $xxx$ (b) shift current densities versus photon energy for a 10~nm wide [110]-oriented GaAs QW. Black solid line: total shift current. Blue dotted line: inter-conduction band contribution $j_{cc'}$. Green dashed line: inter-valence band contribution $j_{vv'}$. Magenta dash-dotted line: inter band contribution $j_{cv}$. Red solid line: without heavy hole-light hole coupling.
The vertical  arrows  highlight  the  photon energies corresponding to interband transitions at ${\mathbf k}_\parallel=0$.}
\label{figure4}
\end{figure}

\subsection{$[110]$-oriented quantum wells}

We consider a GaAs QW of 10~nm width grown along the [110] crystallographic direction using a coordinate system in which $x= [001]$, $y= [1\bar{1}0]$, and $z= [110]$.
In the following we analyze
shift and rectification currents generated by linearly polarized laser pulses.
The laser field has a photon energy of $1.54$~eV, i.e., $80$~meV above the band gap ($E_g=1.46$~eV).
For light polarization along the $y$-axis we obtain the current flowing in the $x$-direction, named $xyy$ current.
The computed time evolution of $xyy$ shift and rectification current densities in the GaAs QW for different laser amplitudes is
displayed in Fig.~\ref{t-evol}(a)-(c) and Fig.~\ref{t-evol}(d)-(e), respectively.
Here, solid (dashed) lines correspond to a non-perturbative complete (perturbative second-order)
solution of the SBE.
When the amplitude of the light field is small the perturbative approximation agrees well with the non-perturbative complete solution, see Fig.~\ref{t-evol}(a) and Fig.~\ref{t-evol}(d) for $E_0=10^4$~V/cm.
In this perturbative regime the shift current follows the envelope of optical pulse intensity while the
rectification current has the shape of its time derivative.
The small difference between the two approaches is due to
contributions from heavy to light hole transitions which are enabled by the
finite bandwidth of the incident pulse
and can be only described in the non-perturbative solution.
When the light field amplitude is large enough such that band filling effects
become relevant
there are strong deviations between the approximate perturbative and the full
non-perturbative results, see Fig.~\ref{t-evol}(b) and (e) for $E_0=2\cdot 10^5$~V/cm and
Fig.~\ref{t-evol}(c), and (f) for $E_0=4\cdot 10^5$~V/cm.
In this regime, the second-order approximation fails to describe the shift current dynamics properly and strongly overestimates the current magnitude.
In particular, the non-perturbative solutions of the SBE result in significantly smaller currents
since phase-space filling limits the strengths of the optical excitations
and due to Rabi flopping in the strongly excited regions of k-space
an oscillatory dynamics is obtained, see solid lines in
Fig.~\ref{t-evol}(b), (c), (e), and (f).
For bulk GaAs the relevance of phase-space filling effects 
in limiting the perturbative $\chi^{(3)}$-scaling of two-color injection currents has been
confirmed recently\cite{stern} and it can therofore be expected that also for shift and
rectification currents non-perturbtive signatures should be observable.

In the following, we focus
on the weak excitation regime ($E_0=10^4$~V/cm)
where the perturbative second-order solution works properly and analyze the dependence
of shift and rectification currents on the other excitation conditions, in particular,
the light polarization and the photon energy.
In Fig.~\ref{figure3} we show the $x$- and $y$-components of the peak value of shift current density (Fig.~\ref{figure3}(a)) and rectification current density (Fig.~\ref{figure3}(b))
as function of $\theta$, where $\theta$ is measured with respect to the $x=[001]$ direction 
and determines the polarization direction of the linearly polarized incident pulse.
The computed current components are well described by a formula which also
arises from a macroscopic symmetry analysis:
$j_x=A+B\cos 2\theta$ and $j_y=C\sin 2\theta$.\cite{bieler06jap,bieler07}
We note that for light polarization parallel to the $x$-axis ($\theta=0^\circ$) 
there is a finite current flowing in the $x$-direction, i.e., $xxx$ current.
Such kind of current is not present in bulk GaAs but does exist in [110]-oriented
GaAs QWs because of the symmetry reduction.

\begin{figure}
\centering
\includegraphics[width=8cm]{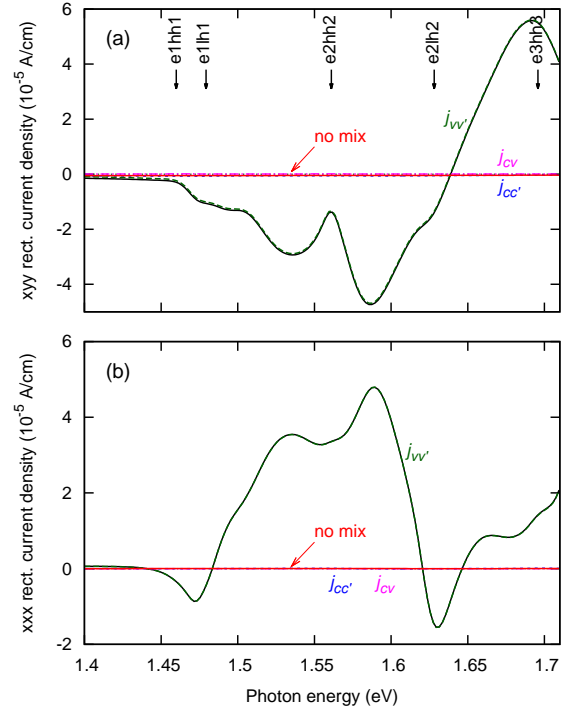}
\caption{(color online) Peak values of the $xyy$ (a) and $xxx$ (b) rectification current densities versus the photon energy
for a 10~nm wide [110]-oriented GaAs QW. The description of the lines is the same as in Fig.~\ref{figure4}.}
\label{figure5}
\end{figure}

The peak values of the $xyy$ ($xxx$) shift current densities versus the photon energy are
presented by the black solid lines in Fig.~\ref{figure4} for $\theta=90^\circ$ ($\theta=0^\circ$), i.e.,
the polarization of the incident field is in $y$-direction ($x$-direction).
With increasing the photon energy nearby and above the bandgap both $xyy$ and $xxx$ shift currents show
a non-monotonic and complex variation.
The $xxx$ current shows a very interesting sign change which corresponds to a reversal
of the current direction at the photon energy of 1.504~eV, see the solid line in Fig.~\ref{figure4}(b).
In order to understand the origin of the complex dependence of shift currents we
separately calculate three different contributions to the net current according to Eq.~(\ref{interbandcurrent}).
Currents originating from inter-conduction band polarizations $p_{cc'}$ ($j_{cc'}$),
inter-valence band polarizations $p_{vv'}$ ($j_{vv'}$), and inter band polarizations $p_{cv}$ ($j_{cv}$) are
evaluated separately and plotted as dotted, dashed, and dash-dotted lines, respectively.
Since $j_{cc'}$ and $j_{cv}$ shown in Fig.~\ref{figure4}(a) follow the step-like increase of the two-dimensional interband density of states
the strong variation of $j_{vv'}$ in both magnitude and direction is responsible for the non-monotonic photon-energy dependence of the net $xyy$ shift current. 
In the case of the $xxx$ current, because of very small contributions of $j_{cc'}$ and $j_{cv}$,
the net shift current is mainly given by $j_{vv'}$.
To obtain non-vanishing $j_{vv'}$ it is necessary that velocity matrix elements ${\mathbf v}_{cv}$, ${\mathbf v}_{cv'}$, and ${\mathbf v}_{vv'}$ are simultaneously non vanishing.
At ${\mathbf k}_\parallel \neq 0$, heavy hole and light hole states are mixed which allows the velocity matrix elements between
all pairs of $c,v$ and $v'$ subbands to have finite and $k$-dependent values.
If we artificially remove the valence band mixing by setting the coupling matrix elements between heavy hole and light hole to zeros
the contribution $j_{vv'}$ vanishes and we obtain a monotonically increasing $xyy$ shift current
and a negligibly small $xxx$ shift current shown as red solid lines in Fig.~\ref{figure4}(a) and Fig.~\ref{figure4}(b), respectively.

The photon energy dependence of $xyy$ and $xxx$ rectification currents is presented in Fig.~\ref{figure5} showing
also a complex, oscillatory variation.
Using the same analysis as for shift currents we find that the dominant contribution to both the $xyy$ and $xxx$ rectification currents
comes from inter-valence band polarizations, i.e., $j_{vv'}$. 
The photon-energy dependence of rectification current is therefore governed by valence band mixing.
Furthermore, the sign change of the optical rectification at the resonance excitation (see Eq.~(\ref{or}) for $\omega=\omega_{cv{\mathbf k}_\parallel}$)
may lead to a direction reversal of the rectification current with increasing photon energy.~\cite{zhang,nastos}

\begin{figure}
\centering
\includegraphics[width=8cm]{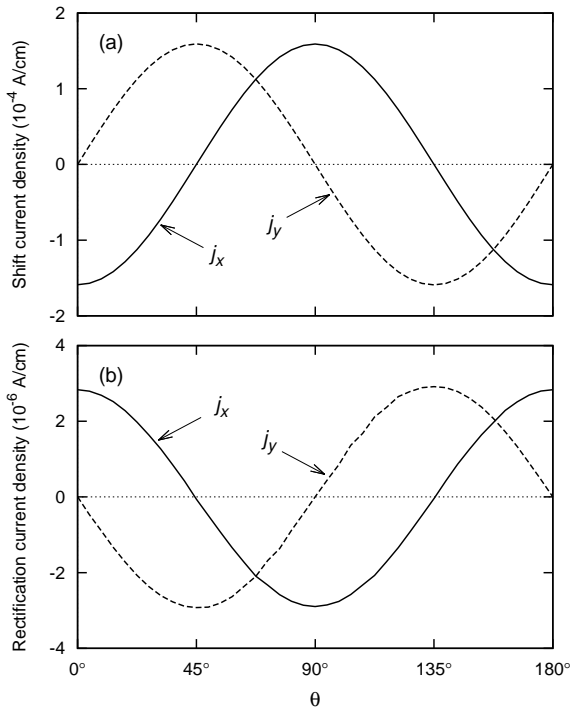}
\caption{Peak value of shift (a) and rectification (b) current densities in $x$- and $y$-directions as function of the light polarization angle $\theta$ in a 10~nm wide [111]-oriented GaAs QW.}
\label{theta111}
\end{figure}

\begin{figure}
\centering
\includegraphics[width=8cm]{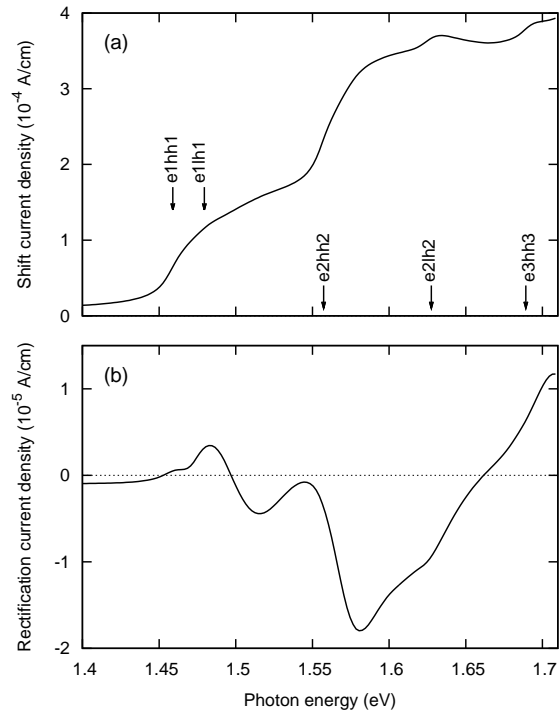}
\caption{Peak value of shift (a) and rectification (b) current densities as function of photon energy in a 10 nm wide [111]-oriented GaAs QW. Vertical arrows show the photon energies corresponding to interband transitions at ${\mathbf k}_\parallel=0$.}
\label{figure7}
\end{figure}

\subsection{$[111]$-oriented quantum wells}

For [111]-oriented QWs
a coordinate system of $x=[11\bar{2}]$, $y=[\bar{1}10]$, and $z=[111]$ is chosen.
We compute the photocurrents in a 10~nm wide GaAs QW using an incident pulse with a peak amplitude of $E_0=10^4$~V/cm.
Figure~\ref{theta111} shows the $x$- and the $y$-components of shift and rectification current densities versus the polarization angle $\theta$ for a photon energy of $1.52$~eV.
The calculated data are fitted well by the formulas: $j_x=A\cos 2\theta$ and $j_y=-A\sin 2\theta$.
The polarization-direction dependence of
the shift and rectification currents in [111]-oriented QW is quite similar
to [111]-oriented bulk GaAs.\cite{zhang}

The photon-energy dependence of the shift and the rectification current densities
is shown in Fig.~\ref{figure7} for the light polarization parallel
to the $y=[\bar{1}10]$ direction ($\theta=90^\circ$).
Similarly to the case of [110]-oriented QWs, also for [111]-oriented QWs
a complex and non-monotonic dependence of 
the shift and the rectification currents on photon energy is obtained.

\section{Comparison to experiments}

Since the shift current involves real carriers its amplitude is typically much larger than the rectification current amplitude under the same excitation conditions, see also Figs.~\ref{figure4},~\ref{figure5}, and \ref{figure7}. Therefore for an analysis of the experiments we only consider shift currents and neglect rectification currents. A comparison between experiments and simulations on injection currents is given in Ref.~\citenum{priyadarshi}.
  
For the shift current experiments we employed a standard free-space terahertz (THz) setup in transmission geometry.~\cite{Mark_4, mark_jap06, mark_ieee08} The samples consist of [110]-oriented GaAs QWs with well widths of 12~nm, 15~nm, and 20~nm. They were excited at normal incidence with a 150~fs laser pulse originating from a Ti:sapphire oscillator with a repetition rate of 76~MHz. The optical peak intensity in front of the samples was 60~MW/cm$^2$ resulting in two-dimensional carrier densities of approximately $2\times 10^{11}$~cm$^{-2}$. The center photon energy of the femtosecond laser could be varied allowing for different excitation conditions. 
The polarization of the initially linearly polarized pump beam was aligned to the $x$ direction of the samples. 

After generation, the shift currents decay on a femtosecond time-scale. Since the emitted electromagnetic radiation is proportional to the time derivative of the current transients, THz radiation is emitted. The THz radiation was collected from the sample’s back surface with an off-axis parabolic mirror, guided to a second off-axis parabolic mirror and focused down onto 1~mm thick [110]-oriented ZnTe crystal. The probe pulse was guided through a small hole in the second off-axis parabolic mirror, collinearly overlapped with the THz pulse and read out the electric field-induced refractive index change of the electro-optic crystal. A silicon wafer and a THz polarizer were placed between the two off-axis parabolic mirrors. The silicon wafer blocked any scattered pump light. The THz polarizer allowed us to detect currents flowing in certain directions in the sample since the polarization of the THz radiation in the far field is parallel to the direction of the current flow. This is a large advantage of THz experiments over experiments based on charge collection at electrodes, since the THz setup is only sensitive to currents flowing in the plane of the sample. In the experiments the THz polarizer was oriented such that only shift currents flowing along the $x$ direction were detected. The bandwidth of the experimental setup was limited by the velocity mismatch between the group velocity of the optical pulse and the phase velocity of the THz pulse in the electro-optic crystal. These restrictions allowed for the detection of frequencies up to approximately 3~THz. All experiments were done at room temperature.  

\begin{figure}
\centering
\includegraphics[width=8cm]{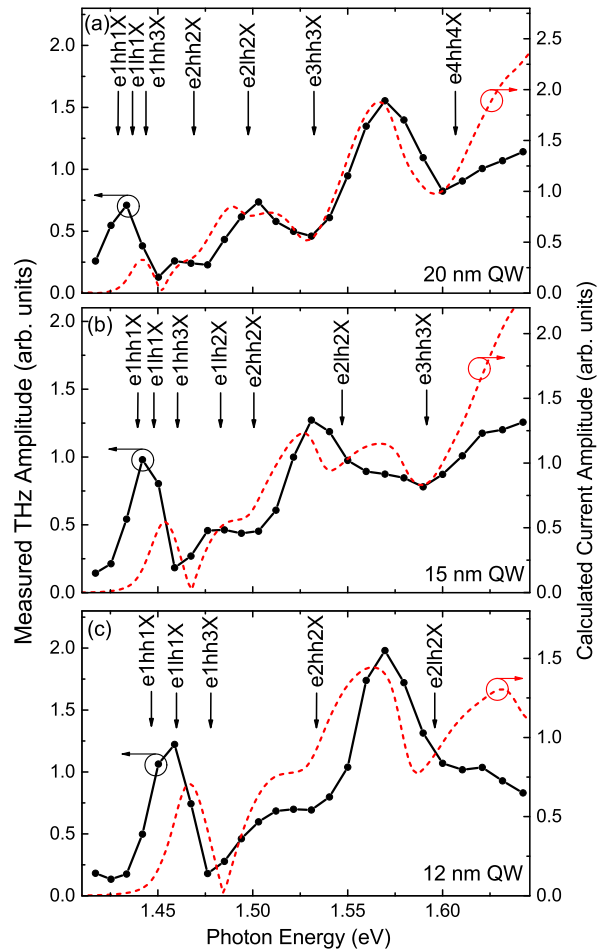}
\caption{(color online) Measured THz amplitudes emitted from shift currents (solid black)
and calculated absolute peak values of shift currents (dashed red)
in x-direction in [110]-oriented GaAs QWs of different well width: (a) 20~nm, (b) 15~nm, and
(c) 12~nm.}
\label{comp}
\end{figure}

In Fig.~\ref{comp} we plot the THz amplitude emitted from the $xxx$ shift currents in the three QW samples versus excitation photon energy. Additionally, the calculated peak magnitude of the $xxx$ shift current is shown.
Starting with the 20~nm QW, we obtain an excellent agreement between theory and experiment in nearly the complete photon energy range. In particular the peaks at 1.5~eV and 1.57~eV and also a zero crossing at 1.45~eV are nicely reproduced. The deviations at small photon energy are most likely because of neglect of Coulomb interaction in the simulations. Moreover the deviations at large photon energies might be the result of improper rescaling of the measured signals. At these photon energies the optical pump and probe power decreased due to the end of the tuning range of our femtosecond laser. 

The deviations between experiment and theory at small and large photon energies also appear for the 15~nm and 12~nm QWs; otherwise the overall agreement between experiment and theory is also good for these QW samples. In general the good agreement between experiment and theory is very important since it (i) validates the theoretical approach and (ii) confirms that the measured signal is mainly caused by the shift of the electron charge during excitation and not by scattering contributions.\cite{sturman}  

\section{Conclusions}

We present an unified microscopic approach which is capable of describing fully dynamically all three kinds of
single-frequency photocurrents in noncentrosymmetric semiconductor QWs, i.e., injection, shift, and rectification currents.
Our approach has been applied to analyze shift and rectification currents generated by linearly polarized incident
pulses in [110]- and [111]-oriented GaAs/AlGaAs QWs.

The dependence of the photocurrents on the polarization direction of the incident laser
pulses has been studied.
For shift and rectification currents we compared 
the non-perturbative solution of the SBE to a second-order approximation.
It is demonstrated that intrinsic nonlinearities arising from phase-space filling effects and Rabi flopping
which are not included in the second-order approximation
lead to a significant difference between the two solutions in the limit of strong excitation intensities ($E_0>10^5$~V/cm). 

Unlike the optical interband absorption which increases in steps with increasing photon energy,
we find an unexpected complex and non-monotonic dependence of shift and rectification currents on
the photon energy in both theory and experiments.
In certain spectral regions both currents may even change their direction.
It has been demonstrated that this dependence is the result of bandmixing of the heavy- and light-hole valence bands.
This excitation frequency dependence of the photocurrents
might be employed for new applications such as opto-electronic converters or modulators. 

\begin{acknowledgments}
Valuable discussions with John E. Sipe are gratefully acknowledged.
This work is supported by the Deutsche Forschungsgemeinschaft (DFG) through the grants
GRK1464/2, ME1916/2-2, and BI1348/1-2.
H.T.D. acknowledges financial support of the Vietnam National Foundation for Science and
Technology Development (NAFOSTED) under the Grant No. 103.62-2012.26.
For providing computing time we thank the PC$^2$ (Paderborn Center for Parallel Computing). 
\end{acknowledgments}


\begin{thebibliography}{50}










\bibitem{belinicher}V. I. Belinicher, E. L. Ivchenko, and B. I. Sturman, Sov. Phys. JETP 56, 359 (1982).
\bibitem{sturman}B. I. Sturman and V. M. Fridkin, {\it The Photovoltaic and Photorefractive Effects in Noncentrosymmetric Materials} (Gordon and Breach, Philadelphia, 1992).
\bibitem{zhang}X.- C. Zhang, Y. Jin, K. Yang, and L.J. Schowalter, Phys. Rev. Lett. {\bf 69}, 2303 (1992).
\bibitem{khurgin}J. B. Khurgin, J. Opt. Soc. Am. B {\bf 13}, 2192 (1996).
\bibitem{kral}P. Kr\'al, J. Phys.: Condens. Matter {\bf 12}, 4851 (2000).
\bibitem{sipe}J. E. Sipe and A. I. Shkrebtii, Phys. Rev. B {\bf 61}, 5337 (2000).
\bibitem{bhat00}R. D. R. Bhat and J. E. Sipe, Phys. Rev. Lett. {\bf 85}, 5432 (2000).
\bibitem{ganichev00}S. D. Ganichev, E. L. Ivchenko, S. N. Danilov, J. Eroms,
W. Wegscheider, D. Weiss, and W. Prettl, Phys. Rev. Lett. {\bf 86}, 4358 (2001).
\bibitem{ganichev02}S. D. Ganichev, U. R\"ossler, W. Prettl, E. L. Ivchenko, V. V. Belkov, R. Neumann, K. Brunner, and G. Abstreiter, Phys. Rev. B {\bf 66}, 075328 (2002).
\bibitem{golub}L. E. Golub, Phys. Rev. B {\bf 67}, 235320 (2003).
\bibitem{cote}D. C\^{o}t\'{e}, N. Laman, and H. M. van Driel, Appl. Phys. Lett. {\bf 80}, 905 (2002).
\bibitem{laman}N. Laman, M. Bieler, and H. M. van Driel, J. Appl. Phys. {\bf 98}, 103507 (2005).
\bibitem{bieler06}M. Bieler, K. Pierz, U. Siegner, and P. Dawson, Phys. Rev. B {\bf 73}, 241312(R) (2006).
\bibitem{bhat05}R. D. R. Bhat, F. Nastos, A. Najmaie, and J. E. Sipe, Phys. Rev. Lett. {\bf 94}, 096603 (2005).
\bibitem{nastos}F. Nastos and J. E. Sipe, Phys. Rev. B {\bf 74}, 035201 (2006).
\bibitem{bieler06jap}M. Bieler, K. Pierz, and U. Siegner, J. Appl. Phys. {\bf 100}, 083710 (2006).
\bibitem{bieler07}M. Bieler, K. Pierz, U. Siegner, and P. Dawson, Phys. Rev. B {\bf 76}, 161304(R) (2007).
\bibitem{n1}J. M. Schleicher, S. M. Harrel, and C. A. Schmuttenmaer, 
J. Appl. Phys. {\bf 105}, 113116 (2009).
\bibitem{priyadarshi}S. Priyadarshi, A. M. Racu, K. Pierz, U. Siegner, M. Bieler, H. T. Duc,
J. F\"orstner, and T. Meier, Phys. Rev. Lett. {\bf 104}, 217401 (2010).
\bibitem{duc10}H. T. Duc, J. F\"orstner, and T. Meier, Phys. Rev. B {\bf 82}, 115316 (2010).
\bibitem{n2}F. Nastos and J. E. Sipe, Phys. Rev. B {\bf 82}, 235204 (2010).
\bibitem{n2a}J. E. Moore and J. Orenstein, 
Phys. Rev. Lett. {\bf 105}, 26805 (2010).
\bibitem{n3}J. W. McIver, D. Hsieh, H. Steinberg, P. Jarillo-Herrero, and N. Gedik,
Nature Nanotechnology {\bf 7}, 96 (2012).
\bibitem{n4}D. A. Bas, K. Vargas-Velez, S. Babakiray, T. A. Johnson, P. Borisov, T. D.
Stanescu, D. Lederman, and A. D. Bristow, 
Appl. Phys. Lett. {\bf 106}, 41109 (2015).

\bibitem{rossler}U. R\"ossler, Solid State Commun. {\bf 49}, 943 (1984).
\bibitem{winkler03}R. Winkler, {\it Spin-orbit Coupling Effects in Two-Dimensional Electron and Hole Systems} (Springer, Berlin, 2003).
\bibitem{elkurdi}M. El kurdi, G. Fishman, S. Sauvage, and P. Boucaud, Phys. Rev. B {\bf 68}, 165333 (2003).
\bibitem{haug}H. Haug and S. W. Koch, {\it Quantum Theory of the Optical and
    Electronic Properties of Semiconductors}, 4th ed. (World Scientific, Singapore, 2004).
\bibitem{pasenow}B. Pasenow, H. T. Duc, T. Meier, S. W. Koch, Solid State Commun. {\bf 145}, 61 (2008).
\bibitem{duc05}H. T. Duc, T. Meier, and S. W. Koch, Phys. Rev. Lett. {\bf 95}, 086606 (2005).
\bibitem{stern}E. Sternemann, T. Jostmeier, C. Ruppert, H. T. Duc, T. Meier, and M. Betz,
Phys. Rev. B {\bf 88}, 165204 (2013).
\bibitem{Mark_4}M. Bieler, K. Pierz, and U. Siegner, Phys. Rev. B {\bf 76},  161304 (2007).
\bibitem{mark_jap06}M. Bieler, K. Pierz, and U. Siegner, J. Appl. Phys. {\bf 100},  83710 (2006).
\bibitem{mark_ieee08}M. Bieler, IEEE Journal of Selected Topics in Quantum Electronics {\bf 14}, 458 (2008).
\end{thebibliography}
\end{document}